# Project X and a Muon Facility at Fermilab

Milorad Popovic

*Fermi National Accelerator Laboratory, Batavia, IL, USA*

**Abstract.** An integrated program is described, starting with muon experiments in the Booster era, continuing with a 2 MW target station, a 4 GeV Neutrino Factory and a 3 TeV Muon Collider, all driven by Project X. This idea provides an integrated approach to the Intensity and Energy Frontiers at Fermilab.

## INTRODUCTION

Project X is a proposed high intensity proton facility intended to support a world-leading program in neutrino and flavor physics over the next two decades at Fermilab while also providing an upgrade path to drive a neutrino factory and/or a muon collider. Project X is an integral part of the Fermilab Roadmap as described in the Fermilab Steering Group Report of August 2007 [1] and of the Intensity Frontier science program described in the P5 report of May 2008 [2]. The primary elements of that research program to be supported by Project X include:

- *A neutrino beam for long baseline neutrino oscillation experiments.* A new 2 megawatt proton source with proton energies between 50 and 120 GeV would produce intense neutrino beams, directed toward a large detector located in a distant underground laboratory.
- *Kaon and muon based precision experiments running simultaneously with the neutrino program.* These could include a world leading muon-to-electron conversion experiment and world leading rare kaon decay experiments.
- *A path toward a muon source for a possible future neutrino factory and, potentially, a muon collider at the Energy Frontier.*

This path requires that the new proton source have significant upgrade potential beyond the initial uses.

This paper suggests that an implementation of Project X based on a CW linac can be part of a continuous synergistic transition from a muon physics program in the "Booster era" to the Neutrino Factory and Muon Collider. It then describes a possible staging of the planned muon experiments and of Project X to provide a graceful transition from the Intensity Frontier to the Energy Frontier.

## FERMILAB ACCELERATOR COMPLEX

At present the Fermilab accelerator complex includes a Proton Source, the Main Injector and Recycler rings, an Antiproton Source, and the Tevatron. The Proton Source is a Linac-Booster complex delivering 8-GeV beam with 15-Hz repetition rate capability. The Recycler ring is an 8-GeV storage ring built out of permanent magnets and presently used for storing antiproton beams. The Main Injector accelerates beam from 8 GeV to 120 GeV every 1.3 seconds; with upgrades, it will be able to accelerate $1.6 \times 10^{14}$ protons per cycle. The Antiproton Source produces antiprotons at a 120-GeV target station and collects and cools them using the Debuncher and Accumulator rings. The Tevatron accelerates protons and p-bars for collisions at 980 GeV.

### Near Future

Soon the Collider program will end and the Antiproton Source will be available for Mu2e and muon g–2 experiments. During the Booster era, both experiments will use beam from the Booster and use the Antiproton Source rings for beam preparation specific to each experiment. Both experiments will be able to run in parallel to the neutrino program from the Main Injector using only protons that cannot be used for the Main Injector program. Using the Recycler as an accumulator ring, the Main Injector will run with a repetition period of 1.33 sec or 20 Booster cycles. Assuming that the Booster will run at a 15-Hz rate, twelve cycles will be used for the Main Injector neutrino program (NOνA), and the other eight will be available for 8-GeV experiments such as Mu2e, g–2, and a Booster-based neutrino program.

ACKNOWLEDGEMENT
Operated by Fermi Research Alliance, LLC under Contract No. DE-AC02-07CH11359 with the United States Department of Energy.

Mu2e, if it runs alone with NOνA, can use all eight of the remaining Booster cycles. In one of the scenarios, two Booster cycles are transferred, using part of the Recycler as a transfer line, and momentum-stacked in the p-bar Accumulator ring. Then this beam is bunched into four bunches and transferred one at a time to the Debuncher ring for slow spill extraction to the experiment.

Similarly, if g–2 runs alone with NOνA, it can use all eight Booster cycles. In that case, as soon as the Recycler has transferred beam to the Main Injector, one Booster batch is transferred into the Recycler, bunched into four bunches. Then every 11 milliseconds a bunch is targeted on the p-bar production target for creation of pions and muons. To provide a long beam path for most of the pions to decay, this beam is transported to the experiment using the AP2 transfer line, a portion of the Debuncher and the AP3 transfer line.

If 30 kW proton beam is of interest to the kaon experiments, the required bunch structure can be created. For kaon experiments, each available Booster cycle can be transferred to the Accumulator ring. This beam can be rebunched in forty bunches to create ~40 nsec bunch spacing. Then this beam can be transferred to the Debuncher ring and slow spilled to the experiment with bunch length of a few hundred picoseconds using the method described in the KOPIO proposal. Like the case of the muon experiments, this beam delivery will not affect the NOνA program.

## Looking ahead

At present the plan is to have each Booster-era experiment at a different location on the Fermilab site, and each of them will need a separate newly built experimental hall. Instead, it is suggested here that the aforementioned 8-GeV experiments can have common beam delivery tunnels and transfer lines. The rest of this note presents one possible option that will house all these experiments in one new building that in the future can become a multi-megawatt target station for a Neutrino Factory and/or a Muon Collider.

To minimize the need for new tunnel, the plan is to extract beam from the Tevatron tunnel in the vicinity of the F0 sector and bend it into the Tevatron infield as shown in Figure 1. It is also assumed that the bend will be part of a future new tunnel that will house accumulation and bunching rings. Then the new tunnel will extend for an additional forty meters and enter a new experimental building. This new building will have the floor space needed for a future 4-MW target station, but initially it will be used to house Mu2e, g–2 and kaon experiments.

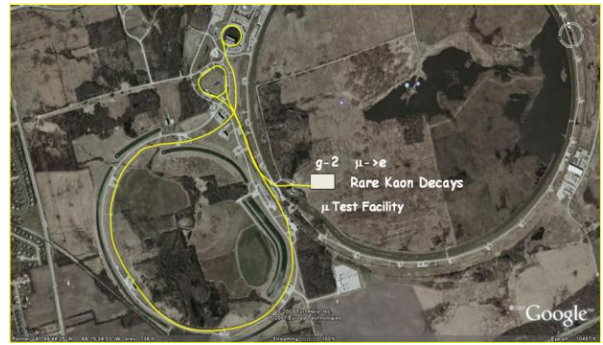

**FIGURE 1.** The proposed location of the new experimental hall.

The location of the new experimental hall will allow delivery of the Booster beam using existing tunnels and beam lines. It will also require construction of a short tunnel and a single building for all three experiments. The size of the building is 50 by 25 meters, defined by the need to be a 4-MW target station in the future as shown in Figure 2.

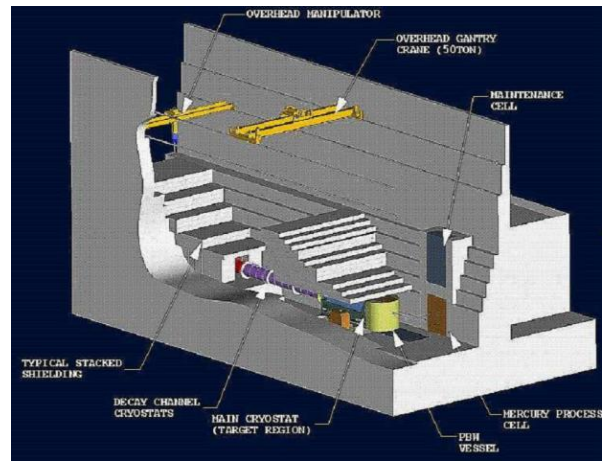

**FIGURE 2.** A conceptual design of the future target station.

To save money initially, the experimental hall will be built just to satisfy the needs of the experiments. After the experiments are finished, adequate shielding will be added for the target station. In the next phase a Muon Capture Channel and a Muon Experimental Area have been added as shown in Figure 3.

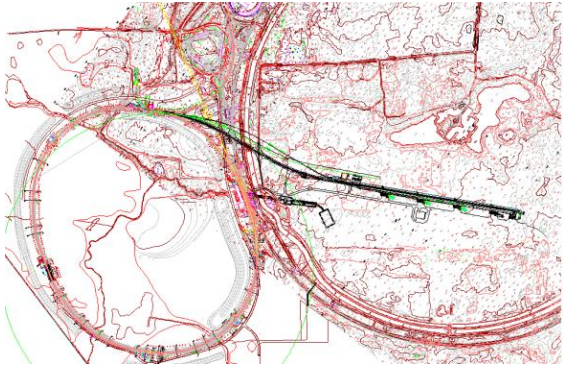

**FIGURE 3.** The location of the Target Station, Capture Channel, and Muon Test Area with respect to the 8-GeV linac.

Once again it is emphasized that in this approach, every phase fully utilizes facilities built for the previous stage. The next phase might be the facilities required for a Low Energy Neutrino Factory as shown in Figure 4.

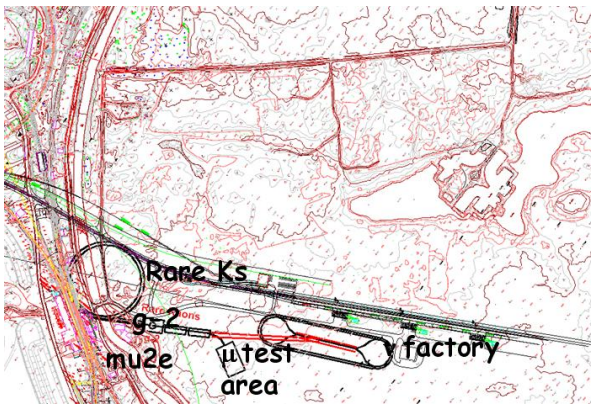

**FIGURE 4.** A proton collection ring, muon accelerators and muon decay ring as part of a Low Energy Neutrino Factory.

## SUMMARY

The concepts presented here emphasize continuity and the need for planning ahead. The approach allows a rich experimental program, maximally utilizes existing facilities and provides a framework for a synergetic and coherent approach to the Intensity and Energy Frontiers.

## ACKNOWLEDGMENTS

Most of the ideas presented here are the result of conversations and work done with Charles Ankenbrandt, currently with Muons, Inc.